\documentclass[twocolumn,showpacs,preprintnumbers,amsmath,amssymb,prl,superscriptaddress]{revtex4}

\usepackage{graphicx}
\usepackage{dcolumn}
\usepackage{bm}
\usepackage{amsmath,amssymb}

\newcommand{\be}{\begin{equation}}
\newcommand{\ee}{\end{equation}}

\newcommand{\beq}{\begin{eqnarray}}
\newcommand{\eeq}{\end{eqnarray}}

\def\eq#1{(\ref{#1})}
\def\H1{\widehat{H}_1}

\begin{document}

\title{Screening in the two-dimensional electron gas with spin-orbit coupling}

\author{M. Pletyukhov}
\affiliation{Institut f\"ur Theoretische Festk\"orperphysik, Universit\"at
Karlsruhe, D-76128 Karlsruhe, Germany}

\author{V. Gritsev}
\affiliation{D\'epartement de Physique, Universit\'e de Fribourg,
CH-1700 Fribourg, Switzerland}
\affiliation{Department of Physics, Harvard University, Cambridge, MA
02138, USA}

\begin{abstract}
We study the interplay between electron-electron interaction and Rashba
spin-orbit coupling in the two-dimensional electron gas. 
Using the random phase approximation we predict new screening properties of 
the electron gas which result in extension of the region of 
particle-hole excitations and in spin-orbit induced suppression of
collective modes. In particular, we observe that the spin-orbit coupling 
gives a finite lifetime to plasmons and longitudinal optical phonons, which 
can be resolved in inelastic Raman scattering. We evaluate the 
experimentally  measurable dynamic structure factor and estimate the range 
of parameters where the described phenomena are mostly pronounced. 
\end{abstract}
\maketitle


Fundamental issues of interaction effects in the two-dimensional electron gas 
(2DEG) are at the center of discussions since the early days of its fabrication
 \cite{AFS}. Quite generally, screening described by the dielectric function 
forms the basis for understanding a variety of static and dynamic many-body 
effects in electron systems \cite{Mahan}. The dielectric function of the 
2DEG was computed a long time ago by F. Stern \cite{Stern} within the 
random phase approximation (RPA). Different quasiparticle and collective 
(plasma) properties deduced from these expressions were confirmed 
experimentally soon after \cite{ATS}. Recent experiments measuring plasmon 
dispersion, retardation effects and damping \cite{Nagao,Kuk,West} 
unambiguously show the importance of correlations between electrons.

More recently, the possibility of manipulating spin in 2DEG by 
nonmagnetic means has generated a lot of activity \cite{ZFS}. 
The key ingredient is  spin-orbit (SO) coupling tunable by an 
applied electric field \cite{Nitta}. A recent example 
for SO-induced phenomena is the spin-Hall effect \cite{SHE}.  

In this context  understanding of the combined effects  
of electron-electron correlations and SO coupling in 2DEG becomes 
vitally important. In the present 
paper we study the RPA dielectric function at finite momenta and frequencies 
treating the SO coupling exactly. We observe two interesting phenomena: an 
extension of the continuum of particle-hole excitations and  SO-induced 
damping of collective modes. New features in the dynamic structure factor 
are predicted, and SO-induced widths of both plasmon and longitudinal optical 
phonon spectra are evaluated. Their values lie in the ranges which are 
experimentally accessible by now in the inelastic Raman scattering 
measurements. We also revisit the derivation of zero-frequency limit of the 
dielectric function and present an analytic result which does not contain an 
anomaly at small momenta \cite{Raikh}.

We consider a 2DEG with SO coupling of the Rashba type \cite{BR} 
described by the single-particle Hamiltonian 
$H =\frac{{\bf k}^{2}}{2m^{*}}+\alpha_{R}{\bf n}(\mbox{\boldmath$\sigma$}
\times {\bf k})$, where  ${\bf n}$ is a unit vector normal to the plane of 
2DEG and $\hbar=1$. The dispersion relation is SO-split into two subbands 
$\epsilon_{{\bf k}}^{\pm} =\frac{k^2}{2 m^*} \pm \alpha_{\mathrm{R}}k$ with
two distinct Fermi momenta $k_{\pm} = k_F \mp m^* \alpha_R \equiv k_F \mp k_R$,
 but with the same Fermi velocity $v_F = k_{F} /m^*$. The effective Coulomb 
interaction  is $V^{eff}_{q \omega}=V_q /\varepsilon_{q \omega}$, where 
$V_q=2\pi e^2/(q\varepsilon_{\infty})$, $\varepsilon_{\infty}$ is the 
(high-frequency) dielectric constant of medium, and the dielectric function 
$\varepsilon_{q \omega} = \varepsilon_{1, q \omega}+ i 
\varepsilon_{2, q \omega}$ describes effects of 
dynamic screening. In RPA, $\varepsilon_{q \omega} =
1-V_q \Pi_{q\omega}$, where $\Pi_{q \omega}$ is the polarization operator.  
In the presence of SO coupling $\Pi_{q \omega} = \Pi^+_{q \omega} 
+ \Pi^-_{q \omega}$ includes contributions from intersubband $(+)$ and 
intrasubband $(-)$ transitions:
\beq\label{bub}
\Pi^{\pm}_{q \omega}\!=\! \lim_{\delta \to 0} \sum_{\mu = \pm}\int 
\frac{d^2 {\bf k}}{(2 \pi)^2} \frac{n_F (\epsilon_{{\bf k}}^{\mu}) -
n_F (\epsilon_{{\bf k}+{\bf q}}^{\pm\mu})}{\omega + i \delta + 
\epsilon^{\mu}_{{\bf k}} - \epsilon_{{\bf k}+{\bf q}}^{\pm\mu}} \,\, 
{{\cal F}}^{\pm}_{{\bf k}, {\bf k}+{\bf q}}, 
\eeq
where the form factors 
${{\cal F}}^{\pm}_{{\bf k}, {\bf k}+{\bf q}} = 
\frac12 [1 \pm \cos (\phi_{{\bf k}} - \phi_{{\bf k}+{\bf q}})]$ 
originate from the rotation to the eigenvector basis.

\begin{figure}[t]
\includegraphics[width =4.25cm]{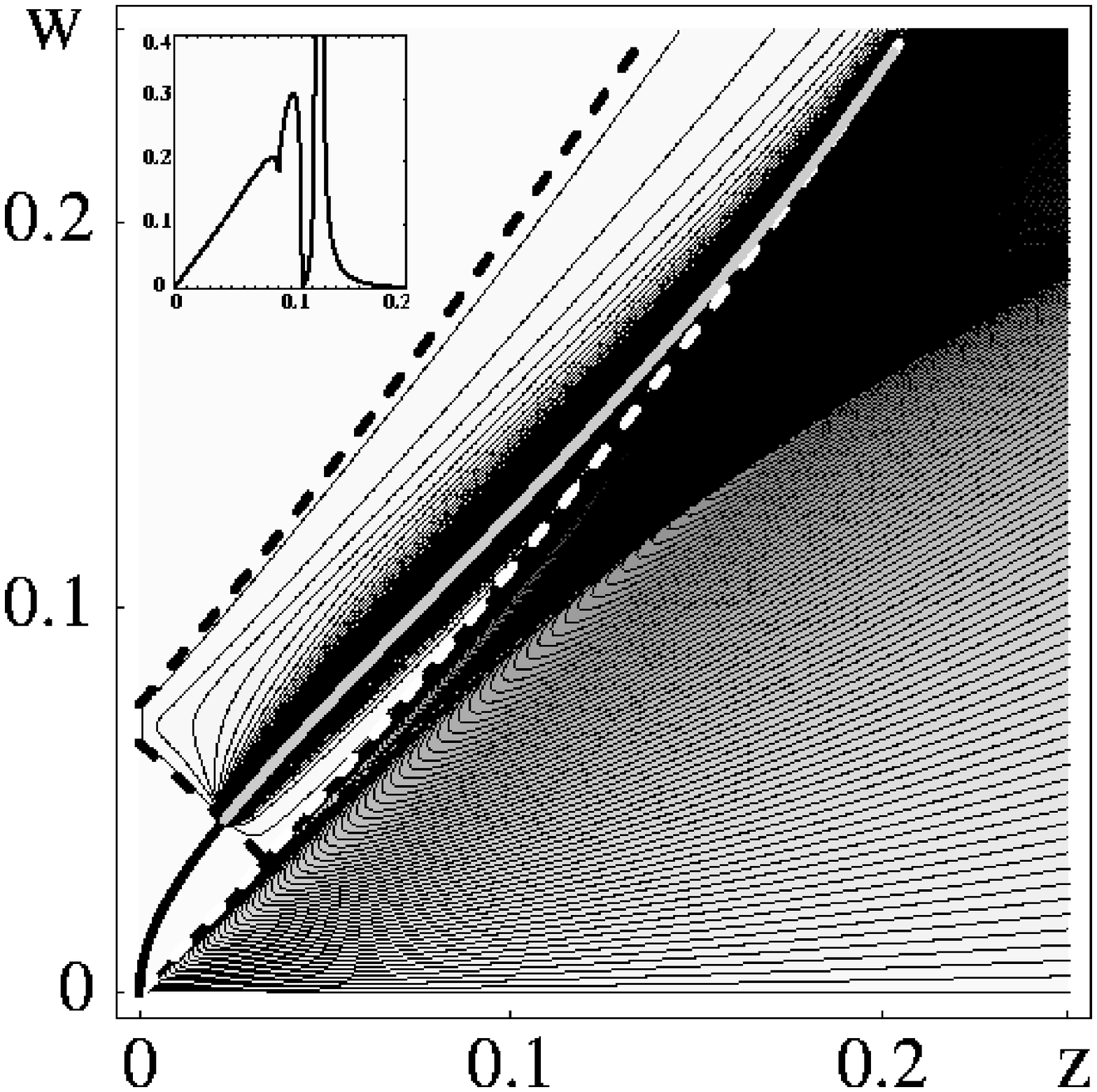}
\includegraphics[width =4.25cm]{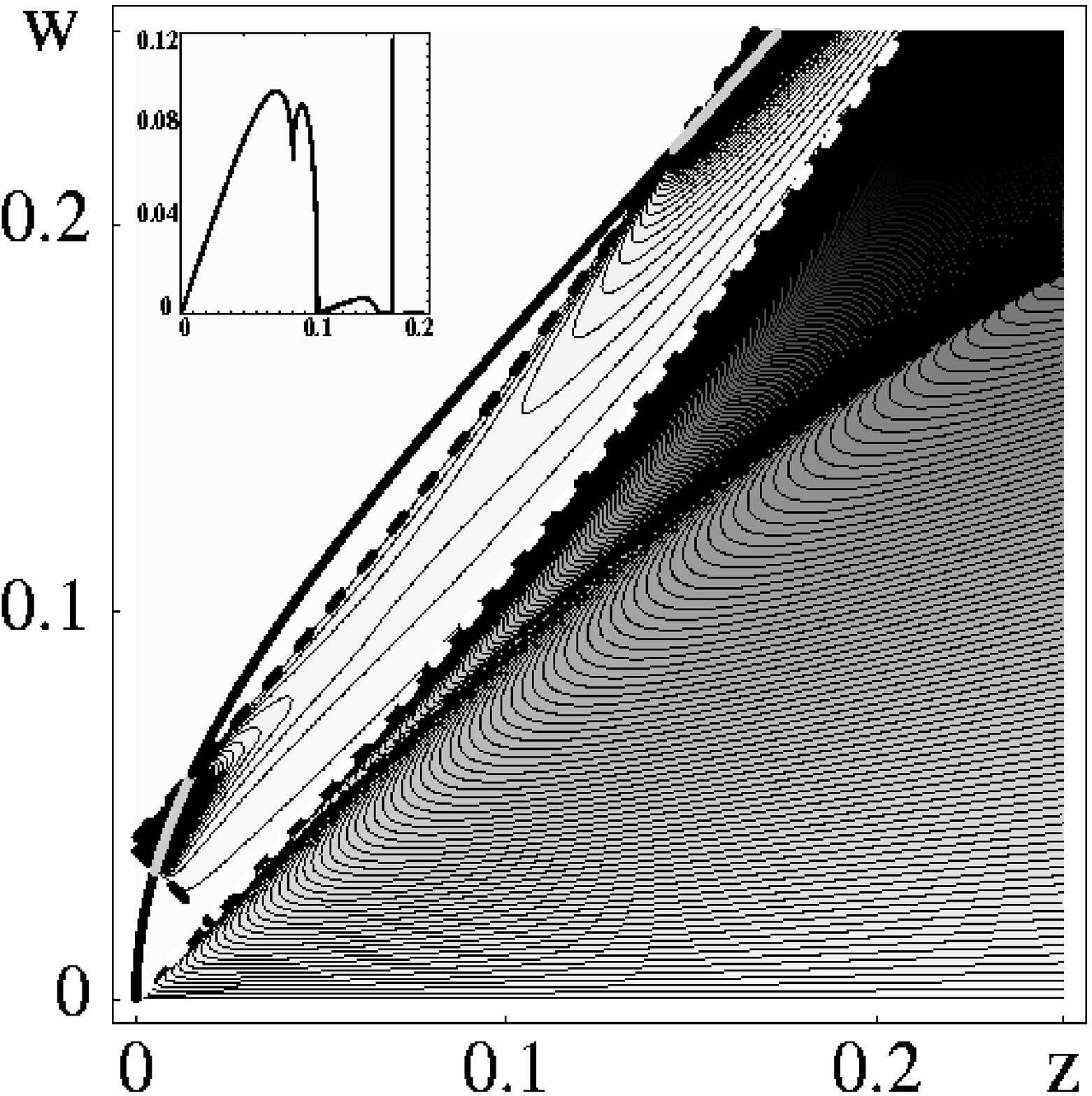}
\caption{Contour plots of $S (z,w)$ showing 
the SO-induced wedge-shaped damping region (bounded by the dashed lines). 
The plasmon mode is depicted by the bold line. 
Insets show the cross-sections $S(z=0.1,w)$ as a function of $w$. 
{\it Left panel:} $y\!=\!0.07$, $r_{s}\!=\!0.2$.
{\it Right panel:} $y\!=\!0.04$, $r_{s}\!=\!0.6$. } 
\label{g0}
\end{figure}

We now discuss our results for $\varepsilon_{q \omega}^{RPA}$. We have derived 
them in the analytic form which will be presented in a subsequent paper.
Here we discuss the main SO-induced effects for the range 
of small $z$ and $w$, and describe how they depend on the SO-coupling strength 
 $y= k_R/k_{F}$ and on the Wigner-Seitz parameter 
$r_s =1/\sqrt{n\pi a_{B}^{*2}}$ which is defined by an electron density $n$ 
and effective Bohr radius $a_{B}^{*}=\varepsilon_{\infty}m_{e}a_{B}/m^{*}$. 
Our main results are compactly 
summarized in Fig.~\ref{g0} which shows the contour plots and cross sections 
of the structure factor 
$S (z, w)=- {\rm Im} \left[1/\varepsilon(z,w) \right]$ for different values 
of  $y$ and $r_s$. The introduced notations  $z=q/2k_{F}$ and 
$w=m^* \omega/2 k_{F}^2$, are, respectively, the dimensionless wave vector 
and frequency.

First of all, we observe an exstension of the continuum of particle-hole 
excitations (or Landau damping region) defined by $\varepsilon_2 \neq 0$. 
A new wedge-shaped region of damping is generated by the SO 
coupling.  It is bounded by two parabolas $-(z-y)^{2}-(z-y) < w <
(z+y)^{2}+(z+y)$ (black-dashed lines in Fig.~\ref{g0}) and attached 
to the boundary $w=z^2+z$ (white-dashed line in Fig.~\ref{g0}) of 
the conventional particle-hole continuum (obtained in the absence of SO coupling according to \cite{Stern}). Another boundary $w = z^2-z$ of the latter transforms into $w = (z-y)^2 - (z-y)$ for nonzero $y$ (this occurs at $z>1$ which are not shown in Fig.~\ref{g0}). The new regions of damping reflect an opened possibility for transitions between SO-split subbands, and $S (z,w)$ strongly depends on $y$ inside these regions. Their boundaries can be determined from a simple consideration of extremes of the denominators in \eq{bub}. 
On the other hand, within the conventional boundaries $w=z^2 \pm z$ 
the function $S (z,w)$ is modified by SO coupling only slightly and can 
be approximated by Stern's expressions \cite{Stern}.

\begin{figure}[b]
\includegraphics[width =8.65cm,angle=0]{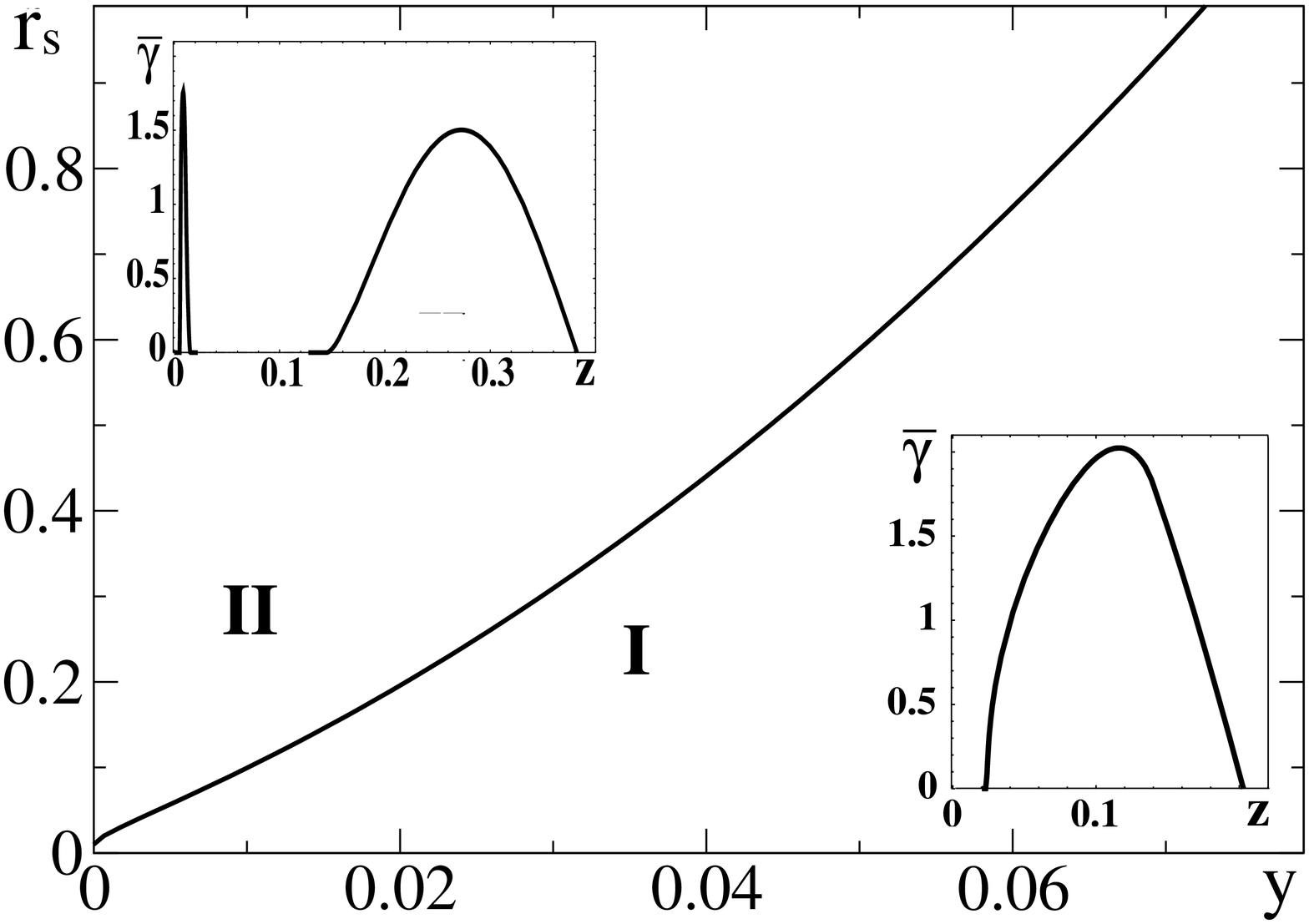}
\caption{The plane $(y,r_{s})$ is divided into the domains I and II where 
the plasmon has one or two undamped pieces, respectively (cf. Fig.~\ref{g0}).  
Insets show SO-induced plasmon width 
$\overline{\gamma}(z)=\gamma (z)\times 10^4$ for the same parameters 
as in Fig.~\ref{g0}.} 
\label{crit}
\end{figure}

The second SO-induced modification is the broadening of 
a plasmon collective mode within the new particle-hole continuum.  
We determine numerically the position of the plasmon spectrum 
from the equation 
$\varepsilon_1 (z,w) = 0$ and depict it in Fig.~\ref{g0} by the bold 
line, in black where $\varepsilon_2  = 0$ 
(undamped plasmon, the structure factor is delta-peaked) and in gray where 
$\varepsilon_2 \neq 0$ (SO-damped plasmon, the structure factor 
has finite height and width). 
Two different cases are possible:
(I) the plasmon enters only once into the SO-induced damping region 
(Fig.~\ref{g0}, left panel); 
(II) it enters twice escaping for a while after the first entrance 
(Fig.~\ref{g0}, right panel). We observe that the position of the plasmon in 
the presence of SO coupling is almost indistinguishable (at least, on the scale of Fig.~\ref{g0}) from the RPA result 
derived in the absence of  SO coupling \cite{Singwi}
\beq\label{plas}
w_{pl}(z)=\frac {z(z+2\tilde{r}_{s})}{2\tilde{r}_{s}}
\sqrt{\frac{4\tilde{r}_{s}^{2}+
4\tilde{r}_{s}z^{3}+z^{4}}{z(z+4\tilde{r}_{s})}} \Theta (z^* -z),
\eeq  
where $\tilde{r}_{s}=r_{s}/\!\sqrt{8}$ and $z^*$ is the real positive root of 
the equation $z^2 (z+ 4\tilde{r}_{s}) = 4 \tilde{r}_{s}^2$. Later we will 
comment more on how small might be the difference between the exact 
SO-modified plasmon dispersion and that given by \eq{plas}. 
At the moment we will use \eq{plas}
in order to calculate critical values of $y$ and $r_s$ which separate 
the cases (I) and (II). The result is represented in 
Fig.~\ref{crit} where the plane $(y, r_s)$  is divided into the respective 
domains I and II. Changing $y$ and $r_s$, we can tune 
the relative position of the new particle-hole region and $w_{pl}(z)$.

In the new region of damping the dynamic structure factor is very well 
approximated by the Lorentzian 
\beq\label{lorentz}
S (z,w)_{SO-damp \,\, pl}=\frac{\alpha (z)}{\pi}  
\frac{\gamma(z)}{(w-w_{pl}(z))^{2}+\gamma^{2} (z)},
\eeq
which describes the SO-damped plasmon with the width $\gamma (z)$. 
The weight factor 
$\alpha (z) = \pi (d\varepsilon_{1}/dw|_{w=w_{pl} (z)})^{-1}$ is practically 
independent of $y$ and given by its value  at $y=0$
\beq\label{plasmon}
\alpha(z)  = \frac{\pi\sqrt{z}[16\tilde{r}_{s}^{4}-z^{4}
(z+4\tilde{r}_{s})^{2}]}{2 \tilde{r}_{s}
\sqrt{(4\tilde{r}_{s}^{2}+4\tilde{r}_{s}z^{3}+z^{4})(z+4\tilde{r}_{s})^{3}}}.
\eeq
At the same time, the SO-induced width $\gamma(z)=\alpha (z)\varepsilon_{2} 
(z , w_{pl} (z))/\pi$ strongly depends on $y$ via $\varepsilon_2 \neq 0$.  
The insets in Fig.~\ref{crit} show $\overline{\gamma} (z)=\gamma 
(z)\times 10^4$ for the values of parameters same as in Fig.~\ref{g0}: 
(I) $y=0.07$, $r_s=0.2$; (II) $y=0.04$, $r_s=0.6$.
The function $\gamma (z)$ is defined for $z \leq z^*$ and vanishes 
in the region where the plasmon is undamped. In the latter case 
$S (z,w)_{pl} = \alpha (z) \delta (w-w_{pl} (z))$.

Comparing positions of peaks in constant-$z$ cross-sections of the exact 
$S (z,w)$ and of the approximate $S (z,w)_{SO-damp \,\,pl}$ \eq{lorentz}, 
we have observed that  the shift of the  plasmon dispersion due to SO coupling 
from $w_{pl} (z)$ \eq{plas} is one or two orders of magnitude smaller than 
$\gamma (z)$ (unless $\gamma (z)=0$), and therefore can hardly be resolved 
experimentally.

It is also worthwhile to compare our exact results with an approximation 
commonly 
used in the literature.  It is based on the formula $\Pi_{q \omega}\approx 
-i\sigma_{\omega}q^2 /(e^2 \omega)$ for small $q$, where $\sigma_{\omega}$ 
is the ac 
conductivity (cf., e.g., Refs.~\cite{MCE,MH,SL}). Such approximation takes 
into account only the transitions with zero momentum transfer, and therefore 
leads to the kinematic extension of the conventional particle-hole continuum 
to a strip $y-y^2 < w <y+y^2$ parallel to the $z$-axis. The lines $w=y \pm y^2$
 coincide with the parabolas $w = \pm (z \pm y)^2 \pm (z\pm y)$ at $z=0$ only, 
i.e. where the above relation between  $\Pi_{q \omega}$ and $\sigma_{\omega}$ 
becomes exact. This approximation can be justified for very small momenta 
(e.g., up to $z \approx 10^{-5}$ for $y=0.01$ and up to $z \approx 10^{-7}$ 
for $y=0.001$). However, it fails to describe properly the behavior of the 
plasmon mode at larger $z$. For example, according to \cite{MCE} the plasmon 
spectrum does not cross the border of their  damping region but rather 
approaches it exponentially from both sides of the strip. This 
approximation also misses the possibility, shown in the left panel of 
Fig.~\ref{g0}, that the plasmon does not emerge undamped for intermediate 
momenta.

An important test of our results is provided by  the Kramers-Kronig relations 
and the sum rules. In particular, we have checked numerically that our 
expressions for $\varepsilon (z,w) -1$ and 
$1/\varepsilon (z,w) -1$ both satisfy the zero-frequency 
Kramers-Kronig relations, which link $\varepsilon_2 (z,w)$ and $S (z,w)$ to 
the static dielectric function $\varepsilon_{1}(z,0) = 
1 - (\tilde{r}_s /z) \cdot (2\pi\Pi_1 (z,0)/m^{*})$. Our exact result for 
zero-frequency polarization operator is
\begin{widetext}
\beq
-\frac{2\pi}{m^*} \Pi_1 (z,0) &=& 2 \,\, \Theta(1-y-z)+\Theta(y-|z-1|)\,\, 
(1+\frac{\pi}{2}  \sin\psi )-2 \,\, \Theta(z-1) \, {\rm arccosh}\,z\, \cos\psi 
\nonumber\\
&+& \sum_{\nu=\pm}\Theta(z-(1+\nu y)) \left( 1+\nu\psi_{\nu}\sin\psi -
\cos\psi_{\nu}-2\cos\psi\ln\frac{1+z\sin(\psi_{\nu}-\nu\psi)}{2\sqrt{2 z}
\cos\frac12\psi_{\nu}\cos\frac12\psi } \right), \label{pzero}
\eeq
\end{widetext}
where $\sin\psi=y/z \leq 1$ and $\sin\psi_{\nu}=(1+\nu y)/z \leq 1$ are 
defined  in the corresponding ranges of $z$ restricted by the unit-step 
function $\Theta$. It is assumed that $y<1/2$ (or $2 k_R < k_F$). For the 
values $z \leq 1-y$ the expression  (\ref{pzero}) yields $-\Pi_1 (z,0) = 
m^{*}/\pi$, the density of states in 2DEG. In deriving $\Pi_1 (z,0)$ 
we first perform the momentum integration in \eq{bub}, and only then take the 
limit $\delta \to 0$. Once this sequence is reversed, there would
arise the extra term  $- \pi \Theta (y-z) \sqrt{(y/z)^2 -1}$ in the rhs of 
\eq{pzero}, which produces an anomaly at $z=y$ (cf. Ref.~\cite{Raikh}).

The effect of SO-induced plasmon broadening described above can be observed 
experimentally. For instance, one  can directly measure $S (z,w)$ by means 
of inelastic light (Raman) scattering (see, e.g., \cite{Raman}). Keeping $z$ 
at a fixed value and varying $y$ and $r_s$, one should observe  different 
patterns with either damped or undamped plasmon, like those shown in the 
insets to Fig.~\ref{g0} where the sections of $S (z,w)$ at constant 
$z=0.1$ are presented. In order to estimate realistic parameters we focus on 
InAs-based 2DEG where the strength of Rashba coupling $\alpha_{R}$ has quite 
large values up to $\sim 3\times 10^{-11}$eVm \cite{large} and prevails over 
the Dresselhaus term \cite{Ganichev}. Most of experiments deal with the 2DEG's 
densities ranging from $n=0.7\times 10^{16}$m$^{-2}$ \cite{Engels} to 
$n=2.4\times 10^{16}$m$^{-2}$ \cite{Nitta}. Taking $\varepsilon_{\infty} 
\approx 12$ and $m^{*} \approx 0.03 m_e$ we obtain the range of 
$y$ from 0.04 to 0.075 and the range of $r_{s}$ from 0.38 to 0.3.  
These values belong to the domain I in Fig.~\ref{crit} where the SO-damping 
of the plasmon is especially important. Choosing the value of 
$\gamma \sim 10^{-4}$, we establish that  
$2\tau_{pl}=\hbar/(4\gamma \epsilon_{F})$ is of the order of 10 ps. 
This can, in principle, be resolved experimentally, since the Raman 
measurements done in II-VI quantum wells \cite{Jus} give a finite lifetime 
of the plasmon mode with the typical value $2\tau_{pl} \sim 0.3$ ps.

Plasmon broadening may be also caused by thermal effects. 
We use the results for 2DEG at finite temperature \cite{Fetter} in order to 
estimate the relative contributions of SO-induced and thermal damping. The 
former dominates over the latter below some characteristic temperature which 
is about 90 K for the parameters of Ref.~\cite{large}. This estimate is made 
at constant $z=0.1$. Note that the thermal damping strongly depends on the 
values of $r_{s}$ and temperature whereas the SO-induced damping has quite 
a weak $r_{s}$-dependence (e.g., compare the insets in Fig.~\ref{crit}). 
This may provide a guide for an experimental separation of these two sources 
of broadening.

Our result for the dynamic structure factor indicates that  the other 
collective mode, longitudinal optical (LO) phonon, will also experience
a pronounced 
SO-induced damping. Its dispersion and lifetime can be obtained from the 
renormalized propagator which is expressed in RPA as \cite{JS}
\beq
D(q,\omega)=\frac{2\omega_{LO}}{\omega^{2}-\omega^{2}_{LO}-2\omega_{LO} M_{q}^{2}\Pi_{q \omega} /\varepsilon_{q\omega}}, 
\label{phonon}
\eeq
where $M_{q}^{2}=\textstyle{\frac{1}{2}}A \omega_{LO} V_{q}$, and $A=1-\varepsilon_{\infty}/\varepsilon_{0}$. On the basis of this expression we can  establish that the LO-phonon width (normalized by $2 k_F^2/m^*$) equals 
\be
\gamma_{LO} (z)= \frac{A S (z,w) w_0^2 \varepsilon_{1}^2}{2 w \varepsilon_{1}^2 + A w_0^2 (\partial \varepsilon_1 / \partial w)} \bigg|_{w =w_{ph} (z)}, 
\ee
where $w_0 = (m^*/2 k_F^2) \,\, \omega_{LO}$ and $w_{ph} (z)$ is the renormalized phonon spectrum. Like in the case of plasmons, we may neglect the dependence of $w_{ph} (z)$ on SO coupling, and find the phonon spectrum from the equation $w^2 = w_0^2 [1 + A (1/ \varepsilon_{1} -1)]$, where $\varepsilon_{1}$ is taken at $y=0$.

The lifetime effects of the LO phonons can be very important for a coupled 
dynamics of carriers and phonons.
An interesting possibility arises when $\omega_{LO} \sim 2 \alpha_R k_F$ (or $w_0 \sim y$). This, for example, holds in InAs, where the value 
$\omega_{LO} \approx 28$ meV gives $w_0 \approx 0.07$.  Changing the Rashba coupling strength $\alpha_R$ by an applied electric field one can manipulate the lifetime of an optical phonon which would result in a modification of transport properties by virtue of the electron-phonon coupling. 

\begin{figure}[b]
\begin{center}
\includegraphics[width=9cm,angle=0]{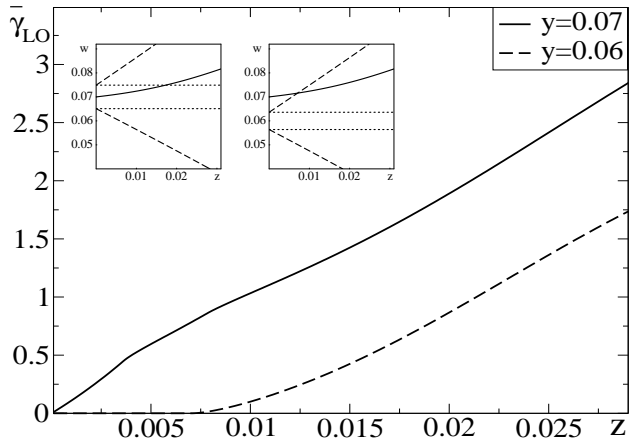}
\caption{
SO-induced damping of the LO phonon  $\overline{\gamma}_{LO} (z) = \gamma_{LO} (z) \times 10^4$ for $w_0=0.07$, $r_s =0.4$, $\varepsilon_{\infty}=12$ and $\varepsilon_0 =15$. Solid and dashed lines correspond to the cases $y=0.07$ ($|w_0 -y|<y^2$) and $y=0.06$ ($|w_0 -y|>y^2$), respectively. Insets show the location of the phonon spectrum (solid line) relative to the SO-induced damping region (dashed lines) and to the strip $y-y^2 < w<y+y^2$ (dotted lines) for $y=0.07$ (left) and $y=0.06$ (right).}
\label{lifephon}
\end{center}
\end{figure}

In Fig.~\ref{lifephon} we show the function $\overline{\gamma}_{LO} (z) = \gamma_{LO} (z) \times 10^4$. In our calculations we used the following parameters: $w_0 =0.07$, $r_s=0.4$, $\varepsilon_{\infty} = 12$ and $\varepsilon_0 = 15$ (for InAs). Choosing the value $\gamma_{LO} = 10^{-4}$ we find that for LO phonons the SO-induced lifetime $2 \tau_{LO}$ is of the order of 10 ps as well. 
We note that the typical lifetime for the LO phonons  measured in AlAs and GaAs by the Raman spectroscopy is of the same order \cite{LOphonon}. 
One can observe that for $y=0.07$ ($|w_0-y| < y^2$)  the damping is always 
finite (solid line), while for $y=0.06$ ($|w_0-y| > y^2$) it becomes nonzero only after some value $z \approx 0.007$ (dashed line). In the two insets we show the relative location of the phonon spectrum (solid line) and of the SO-induced damping region (dashed lines). We also depict the boundaries of the strip $w = y\pm y^2$ (dotted lines), and one can see that for $y=0.07$ the phonon spectrum leaves it at the value $z \approx 0.017$ (left inset), while for $y=0.06$ the phonon spectrum never gets inside the strip (right inset).

In conclusion, we have calculated the dielectric function of 2DEG with Rashba 
spin-orbit coupling in RPA. We have found that the spin-orbit coupling leads 
to a number of new screening properties: the region of particle-hole 
excitations is extended, and the plasmon mode experiences additional 
broadening. The same mechanism leads to the generation of the SO-induced 
lifetime for the longitudinal optical phonons. Thus, spin-orbit coupling tends 
to suppress collective excitations. At the same time, it 
does not affect much the position of the collective mode dispersions. 
Comparison of our theoretical estimates for SO-induced lifetimes with the 
values obtained in the recent Raman scattering measurements suggests that the 
effects described here can be observed experimentally.

We are grateful to Dionys Baeriswyl, Gerd Sch\"{o}n and Emmanuel Rashba  
for valuable discussions. M.P. was supported by the DFG Center for Functional 
Nanostructures at the University of Karlsruhe. V.G. was supported by the Swiss
National Science Foundation through grant Nr.20-68047.02.

\end{document}